# Single-mode sapphire fiber Bragg grating

**MOHAN WANG,**[1] **PATRICK S. SALTER,**[1] **FRANK P. PAYNE,**[1] **ADRIAN SHIPLEY,**[2] **STEPHEN M. MORRIS,**[1] **MARTIN J. BOOTH,**[1] **AND JULIAN A. J. FELLS,**[1,*]

[1]*Department of Engineering Science, University of Oxford, Parks Road, Oxford, OX1 3PJ, UK*
[2]*Rolls-Royce Plc, Derwent Building, 5000 Solihull Parkway, Birmingham Business Park, Birmingham, B37 7YP, UK*
**julian.fells@eng.ox.ac.uk*

**Abstract:** We present here the inscription of single-mode waveguides with Bragg gratings in sapphire. The waveguide Bragg gratings have a novel multi-layer depressed cladding design in the 1550 nm telecommunications waveband. The Bragg gratings have a narrow bandwidth (<0.5 nm) and have survived annealing at 1000°C. The structures are inscribed with femtosecond laser direct writing, using adaptive beam shaping with a non-immersion objective. A single-mode sapphire fiber Bragg grating is created by writing a waveguide with a Bragg grating within a 425 μm diameter sapphire optical fiber, providing significant potential for accurate remote sensing in ultra-extreme environments.

## 1. INTRODUCTION

Fiber Bragg grating (FBG) sensors are widely used for remote monitoring of critical infrastructure applications, because of their ability to measure a range of parameters whilst withstanding extreme environments. Silica optical fiber is generally used, but the operational temperature range of silica FBGs is constrained to significantly below 1000°C. However, sapphire optical fiber has a high melting temperature of ~2054°C, which makes it a promising candidate to extend the limit of current extreme environment sensing applications. For example, gas turbines in aero engines operate at temperatures in excess of 1300°C and the ability to monitor the temperature distribution within them could enable significant improvements in efficiency and emission reduction. Sapphire is also radiation-hard, allowing measurements in nuclear reactors and avoiding radiation-darkening in space applications.

Sapphire optical fiber is a single-crystal, with a large core diameter, no cladding, and a very high refractive index (1.746 at 1550 nm). Sapphire fiber is therefore intrinsically multimode, with over 20,000 modes present in commercially available 75 μm diameter fiber at 1550 nm [1–3]. Femtosecond laser direct writing has been widely used to inscribe FBGs that can withstand extreme environments [4]. For example, sapphire FBGs have been fabricated using the phase-mask [5,6], point-by-point [7], line-by-line [8], and filament-by-filament [9] approaches. However, such gratings have an extremely wide bandwidth (typically ~20 nm) because each mode has a different effective refractive index and consequently a different Bragg resonance for a uniform pitch grating. Furthermore, coupling between modes results in the power distribution fluctuating across the spectrum. Very recently, the bandwidth of a sapphire FBG has been reduced to 1.56 nm using a helical grating structure [10]. However, the fiber is still intrinsically multimode, giving rise to a spectral shift as different modes are excited. The multimode behavior of the sapphire fiber has therefore prevented the widespread adoption of sapphire FBGs for commercial sensing applications.

A single-mode sapphire fiber is therefore needed to allow accurate FBG sensing. Moreover, a single mode sapphire fiber would also enable many alternative techniques, such as scattering based distributed sensing, interferometric sensing and transmission of coherent signals within extreme environments. There have been various attempts to fabricate single-mode sapphire fibers. One method is to reduce the diameter of the sapphire fiber using chemical or physical pre-processing, to the extent that it can only support a restricted number of modes [11,12]. However, the resulting few-mode fiber is only 9 μm in diameter and therefore mechanically vulnerable. A few-mode fiber has also been demonstrated using irradiation of Li-6 enriched lithium carbonate to form a cladding within the fiber [13], but there was failure above 300°C. Alternatively, a few-mode FBG inside a sapphire derived fiber with sapphire core and silica-based cladding has been demonstrated [14]. However, the operating temperature range is constrained to less than 1000°C by the temperature limit of the silica cladding material. Recently, by utilizing the three-dimensional micromachining capability of femtosecond laser processing, single-mode waveguides have been demonstrated in commercially available sapphire bulk using a depressed cladding waveguide (DCW) in the mid-infrared at 2850 nm [15]. However, fabricating single-mode DCWs at telecommunications wavelengths (e.g. 1550 nm), where commodity optical components are available, is significantly more challenging, as the waveguide dimensions required are so much smaller.

We previously proposed that a multimode sapphire fiber could be micromachined to inscribe a single-mode waveguide along its length containing FBGs [16]. In this paper, we fabricate novel single-mode waveguides in sapphire in the 1550 nm waveband and inscribe Bragg gratings within these waveguides. Furthermore, we inscribe these structures within sapphire optical fiber to form single-mode sapphire fiber Bragg gratings.

## 2. LASER FABRICATION SYSTEM AND CALIBRATION

The fabrication system is illustrated in Fig. 1. A regenerative femtosecond laser system (Light Conversion, Pharos SP-06-1000-PP) is used at a second harmonic generation wavelength of 515 nm. The pulse duration was 170 fs and the output beam was linearly polarized. Pulse energies between 10 to 300 nJ were selected by adjusting a half-waveplate prior to a polarizer. The repetition rate was tuned between 10 kHz and 1 MHz. The laser beam was expanded using a telescope and directed onto a liquid crystal spatial light modulator (SLM, Hamamatsu X10468). The phase image from the SLM was imaged to the pupil plane of an air objective (40×, 0.75NA) using a 4-f imaging telescope. Phase correction was adaptively applied to the SLM, to optimize the laser focal spot and mitigate the aberration due to the refractive index mismatch between the sapphire and air [17].

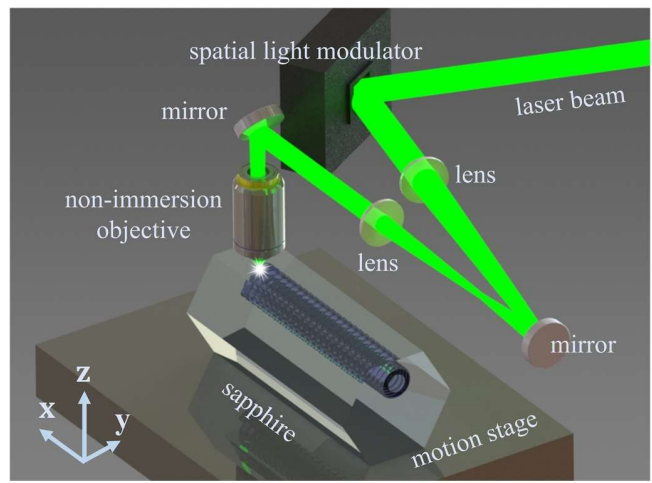

Fig. 1. The femtosecond laser fabrication system.

The sample was mounted on a three-axis motion stage (*x*, *y*: Aerotech ABL10100L and *z*: ANT95-3-V) and translated along the motion stage *x* and *y* axes at a speed between 0.1 to 25 mm/s. Laser modifications were introduced along the *c*-axis of 10×10×1 mm *M*-plane sapphire substrates (Pi-KEM Limited) at a depth of 200 µm below the surface, with the polarization direction parallel to the writing direction. A range of values for pulse energy, repetition rate, and writing speed were used to write single-tracks within the bulk sapphire, in order to establish the parameter window for laser-crystal modification, following a procedure similar to that described in [15]. Fig. 2(a) and 2(b) show the top and cross-sectional views respectively, of the femtosecond laser inscribed tracks at a fixed repetition rate of 1 MHz and a stage translation speed of 11 mm/s. The tracks are for increasing laser pulse energies between 30 nJ to 175 nJ, from left to right. The threshold energy was found to be ~15 nJ, below which the pulse energy became insufficient to generate nonlinear absorption inside the sapphire crystal. The tracks appeared black under the reflection microscope, likely because of scattering due to the formation of subwavelength nanograting structures [18,19].

Fig. 2(c) summarizes the dimensions of the single-tracks written at different repetition rates and pulse energies, at a writing speed of 11 mm/s. The femtosecond laser inscribed track exhibited an asymmetrical 'carrot-shape', related to the energy density distribution at the laser focus and various nonlinear absorption mechanisms [20]. We measured the height and width as the longest dimension along the vertical (laser beam propagation direction) and horizontal (motion stage *y* direction) axes. It was found that the height of the single-track increased monotonically with increasing laser pulse energy and repetition rate, from 3.85 µm (at a pulse energy of 34 nJ and repetition rate of 10 kHz) to 26.36 µm (at 128 nJ and 1 MHz). It was also observed that the width of each single-track was consistently below 3 µm, regardless of the fabrication parameters.

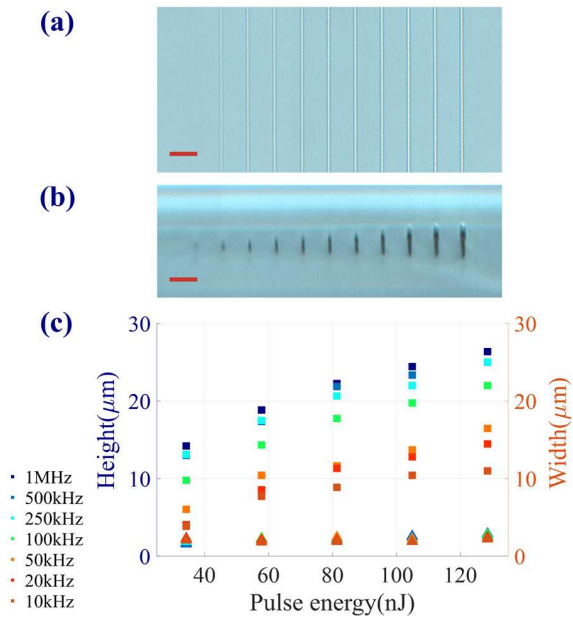

Fig 2. Top-view (a) and cross-sectional view (b) of the laser-induced single-tracks using a repetition rate of 1 MHz and a scan speed of 11 mm/s, at increasing laser pulse energies, measured using a reflection microscope; the red bar indicates a length of 20 µm; (c) graph of width (triangle) and height (square) of the femtosecond laser-written single-tracks for different pulse energies, at repetition rates between 10 kHz and 1 MHz.

## 3. SINGLE-MODE WAVEGUIDES IN PLANAR SAPPHIRE

The regions of the sapphire directly exposed to the laser pulses undergo a decrease in refractive index. A DCW is formed by leaving the core unexposed and exposing the material surrounding the core to reduce its refractive index. Using the data from the single-tracks, a single-mode DCW was designed. The modification dimensions for 30 nJ laser pulse energy, 1 MHz repetition rate and a scan speed of 11 mm/s were used. The design consisted of thirty-four overlapped single-tracks, following an elliptical shape as shown in Fig. 3(a). The number of thirty-four was chosen empirically, to increase the overlap of individual tracks, whilst mitigating the formation of cracks. Fig. 3(b) shows the cross-sectional view of the fabricated DCW in *M*-plane sapphire bulk. The *c*-axis was along the waveguide length to be consistent with sapphire fiber. The outer dimensions for the major (vertical) and minor (horizontal) axes were measured to be ~32 μm and ~17 μm, respectively, while the corresponding values for the inner dimensions were ~15 μm and ~10 μm, respectively.

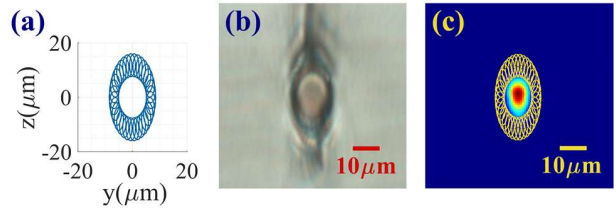

Fig. 3. (a) the DCW design; (b) microscope image of the fabricated DCW from the side facet; and (c) the measured mode profile at 1550 nm with the waveguide design superimposed on top.

To characterize the waveguiding performance of the DCW, a tunable laser source (ID Photonics, CoBrite) was connected to a standard single-mode fiber with a cleaved end (Corning SMF28e+) and butt-coupled to the sapphire end facet. The near-field output mode field was imaged using an 80× microscope objective (Olympus ULWDMSPlan80) and lens (f = 100 mm achromatic doublet) onto an InGaAs camera (Hamamatsu C14041-10U). Fig. 3(c) shows the measured mode profile at 1550 nm, superimposed with the DCW design.

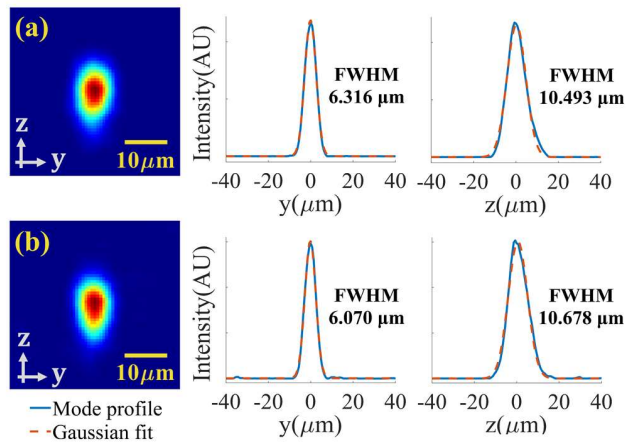

Fig. 4. The experimentally measured guided mode profiles of a DCW at 1550 nm for TE (a) and TM (b) modes. Adjacent are their respective mode fields along the axis (blue straight line) together with a Gaussian fit (orange dashed line).

A polarizer plate was placed in front of the camera and the polarization of the input light was adjusted with a manual controller. Fig. 4 shows the mode profiles of the DCW. The full-width-half-maximum (FWHM) for the TE mode (polarized along the *y* direction) was measured to be 6.32 μm horizontally and 10.49 μm vertically [Fig. 4(a)], while these values were 6.07 μm and 10.68 μm, respectively, for the TM mode (polarized along the *z* direction) [Fig. 4(b)]. The total insertion loss was measured to be between 10.96 to 11.02 dB, (polarization dependent). The waveguides therefore exhibit extremely low polarization dependent loss. The loss includes the Fresnel reflection loss, coupling loss due to the mode mismatch between single-mode fiber and DCW, propagation loss for the 1 cm waveguide, and an extra light diffraction loss due to an edge chamfer leading to a gap of ~250 μm between the DCW and the bulk end facet.

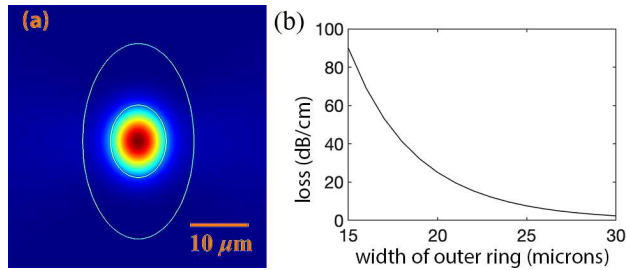

Fig. 5. (a) Mode intensity profile of the DCW computed in FIMMWAVE™ and (b) mode loss as a function of outer ring minor axis width for a fixed aspect ratio of 1.75:1, computed in FIMMWAVE™.

The measured core dimensions and mode widths were used to estimate the laser-induced refractive index change, assuming the core refractive index is unchanged. A two-dimensional Gaussian was fitted to the measured mode widths using a variational best fit. Using a method in [21], the refractive index modification was estimated to be $\Delta n = -2 \times 10^{-3}$. Simulation results using FIMMWAVE™ (Photon Design Ltd.) are shown in Fig. 5 for the waveguide, confirming it to be single-mode. The next higher mode ($E_{y11}$) is predicted to occur at a core size of 11.66 × 15.16 μm. There is therefore a good margin for single-mode behavior. The waveguide loss is dominated by mode-leakage due to the finite cladding width as shown in Fig. 5(b) and could be reduced by increasing this width. The simulated losses appear to be higher than observed, which is likely due to the uncertainty in estimating the extent of the modified region.

The loss is dominated by the minor axis, so an optimum design would have a circular cladding. Bérubé *et al*. achieved a loss of 0.37 dBcm$^{-1}$ in the mid-infrared (2850 nm) [15]. If similar losses could be achieved in sapphire fiber at telecommunication wavelengths, this could allow fiber lengths of 25 cm or more to be used in reflection. Such lengths would be more than sufficient for many applications, for example passing through the turbine casing of an aero engine. Using simulations based on the inferred laser-inscribed refractive index change, the required cladding diameter to reduce the waveguide leakage loss to 0.1 dBcm$^{-1}$ was found to be 49 μm. Reducing the waveguide loss to this figure would allow 1-m lengths of fiber to be used, enabling many new applications.

## 4. SINGLE-MODE WAVEGUIDE BRAGG GRATINGS IN PLANAR SAPPHIRE

Single-mode waveguide Bragg gratings (WBG) have been previously demonstrated in various crystals and different waveguide geometries. A WBG has been demonstrated using a DCW and a Bragg grating in LiNbO$_3$ for electro-optical [22] and quasi-phase matching devices [23]. However, these would not be suitable for high-temperature applications [20,24]. An alternative candidate for high-temperature stability is the dual-line waveguide, which we previously demonstrated on a diamond substrate [25]. However, a dual-line waveguide would not be suitable for transferring to an optical fiber as there would be significant bend loss.

A new approach was therefore needed to fabricate WBGs suitable for sapphire fibers requiring high temperature stability. We developed a three-step process as follows: Step 1: The bottom layers of the multi-layer DCW were first fabricated using a transversal writing method [Fig. 6(a1)]. Step 2: Each individual period of the WBG was written as a ring using a longitudinal writing method [Fig. 6(a2)]. Step 3: The top layers were written using the same parameters as the bottom layer [Fig. 6(a3)]. The whole process is illustrated in Fig. 6(b), with the 3-D design of each fabrication step displayed from left to right. The WBG fabrication method is similar to one employed for bulk glass material [26], where the refractive index modulation is formed by utilizing the intrinsic shape of the Gaussian beam along the beam propagation direction. Fig. 6(c) shows the top-view of a fabricated WBG structure.

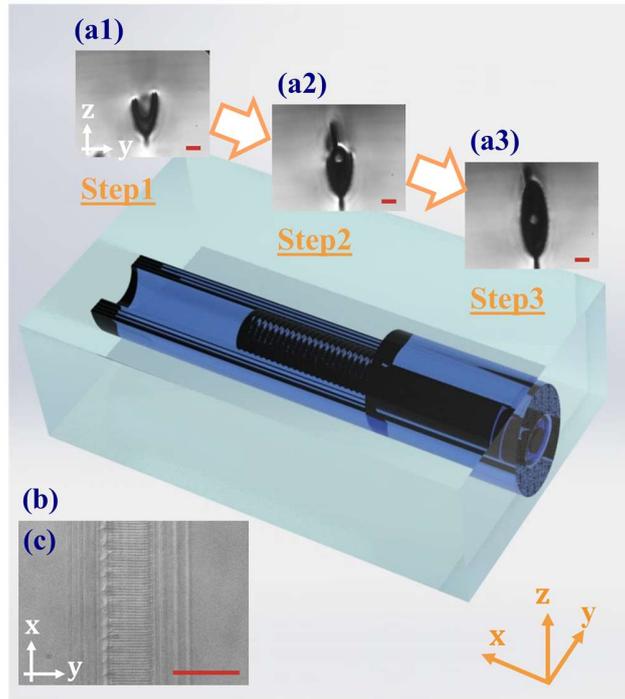

Fig. 6. (a) cross-sectional views for Step 1(a1), Step 2 (a2), and Step 3 (a3) during the three-step fabrication process, (b) diagram of the three-step multi-layer WBG fabrication process, and (c) the top-view of a fabricated multi-layer second-order WBG (Step 3). The red bars indicate a length of 20 μm.

Waveguide Bragg gratings were fabricated following the three-step process to the dimensions of the DCW shown in Fig. 7(a). This DCW had a further 3 outer layers compared with that in Fig. 3(a), in order to reduce the waveguide loss. We fabricated a second-order grating with a period of 887.76 nm. For Steps 1 and 3, a 22 nJ pulse energy, 1 MHz repetition rate, and 11 mm/s scan speed were used, while for Step 2, a speed of 0.1 mm/s and 30-nJ pulse energy were used. The cross-sectional view of the structure is shown in Fig. 7(b). It is noted that cracks in the crystal occur both along the beam propagation and sample translation directions during the fabrication process, particularly at the top and bottom of the DCW structure [Fig. 3(b) and Fig. 6(a)]. While it was found experimentally that these cracks can be mitigated by lowering the laser pulse energy or reducing the overlapping of neighboring single tracks, such approaches also result in either a decrease in the refractive index modification or sacrifice the homogeneity in the DCW region. Though the cracks would distort beam focus at the nearby region and cause inconsistency during fabrication, they resided outside the guiding area and did not incur significant degradation for the guiding behavior of our multilayer structure.

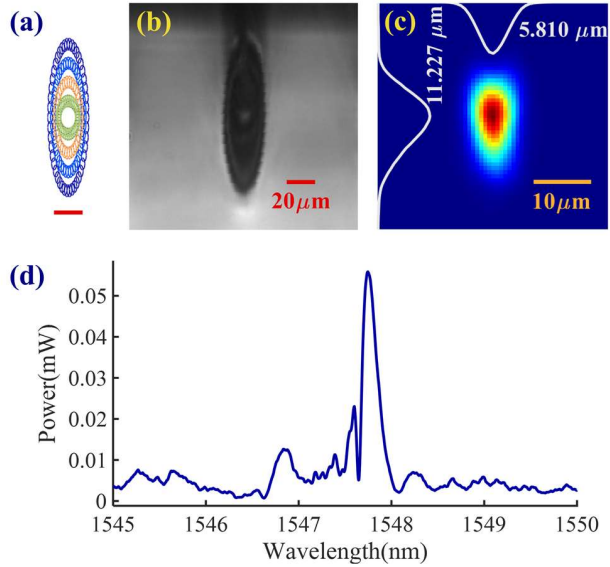

Fig. 7. (a) The design schematic of the multi-layer WBG, the red bar indicates a length of 20 μm, (b) the cross-sectional view of the fabricated sample, measured using a transmission microscope, (c) experimentally measured guided mode at 1550 nm, with the mode profile along the horizontal and vertical central axis plotted at the top and the left axes, and (d) the experimentally measured reflection spectrum.

After fabrication, the guided mode field was measured using the same method as the DCW, shown in Fig. 7(c). The total insertion loss of the 1 cm WBG was measured to be between 6.85 to 9.84 dB at 1550 nm depending on the input light polarization. The spectral performance of the WBG was characterized using a tunable laser source and photodetector system (Agilent 8164A), with a 3-dB coupler. Fig. 7(d) shows the reflection spectrum at room temperature from 1545 to 1550 nm in 2 pm steps. The WBG has a Bragg wavelength, $\lambda_B$ = 1547.75 nm. We believe that the small additional peak at 1547.6 nm is actually the result of interference fringes caused by a Fabry-Perot cavity formed by reflection at the facets. In a commercial device, an antireflection coating could be applied to the facets to mitigate this effect. Whilst the fringes make it difficult to accurately determine the WBG bandwidth, the FWHM would appear <0.5 nm. The mean effective refractive index for the waveguide structure was calculated to be $n_{\text{eff}}$ = 1.7437, from the Bragg equation $m\lambda_B = 2n_{eff}\Lambda$, for order $m$ and pitch $\Lambda$.

To verify the high-temperature performance, a three-layer sapphire WBG written with the same fabrication parameters described above was annealed inside a box furnace. The temperature was raised up to 1000°C over 7 hours, kept at 1000°C for an hour, then decreased to room temperature over 7 hours. Fig. 8 shows the reflection spectrum of the WBG before and after annealing. The wavelength shifted from 1547.93 nm to 1548.43 nm, equivalent to an increase in the mean effective refractive index of the WBG from 1.7439 to 1.7444 after annealing.

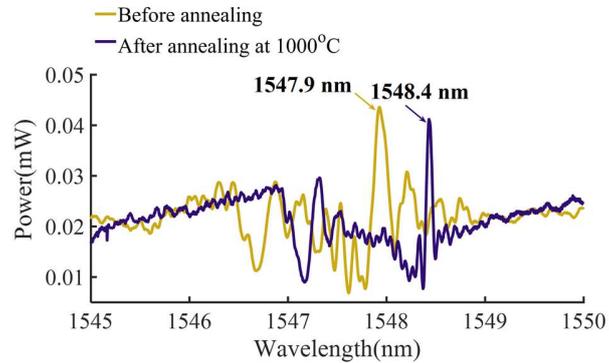

Fig. 8. The reflection spectrum of a WBG before and after annealing.

## 5. SINGLE-MODE SAPPHIRE FIBER BRAGG GRATING

The fabrication technique described above was applied to create single-mode FBGs in sapphire fibers. A commercial sapphire fiber with a diameter of 425 μm and the *c*-axis along its length was used (Photran). Sapphire fiber usually has a hexagonal shape from the manufacturing process. However, the surfaces of the six sides are not perfectly flat and have curvatures towards the radial direction. Using the motion stage and the imaging system, the fiber surfaces were profiled, and an arc was fit to each surface of the sapphire fiber, to calculate an approximate radius of curvature. The aberration resulting from this curvature was calculated and compensated using a method we developed for silica optical fibers [27]. The required phase compensation to focus the laser beam at different positions within the sapphire fiber cross-section was calculated and decomposed into Zernike coefficients. This phase correction was applied to the SLM dynamically in real-time during fabrication.

It was observed that the dimensions of each single track written in the fiber was much smaller than those in the bulk sapphire at the same parameter settings, which was likely due to incomplete correction of the aberration at the cylindrical fiber surface. The waveguide design was adjusted to take into account the change in dimensions, to create a single-mode guiding area with surrounding outer layers. A second-order FBG was inscribed in the fiber, surrounded by two outer layers at a depth between 50 and 100 μm below the fiber top surface. The geometry was carefully designed such that the FBG in the center had a guiding area dimension consistent with the single-mode DCW dimensions we found in the modeling and the WBG on the planar sapphire. The same laser pulse energy and repetition rate as that of the planar waveguide were used, while the translation speed was adjusted to 1-mm/s for Step 1 and 3. A 1-cm long FBG was fabricated, shown in Fig. 9(a-b). Crack formation could be observed along the FBG at the top.

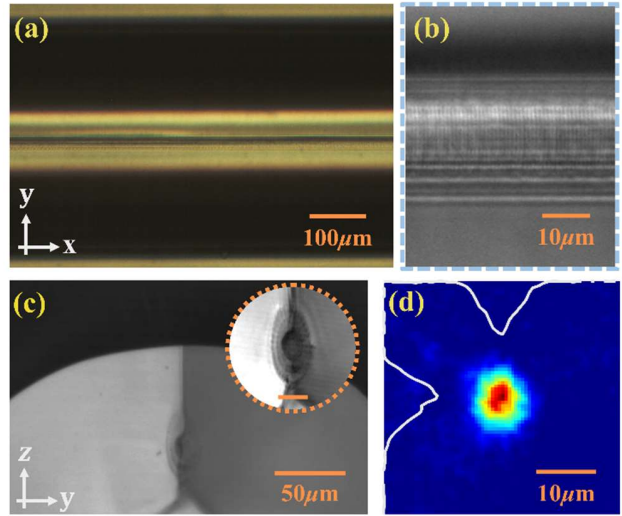

Fig. 9. (a) Top-view microscope image of a sapphire fiber with an FBG inscribed in the center, (b) a magnified view of (a), (c) microscope image of the end-facet, with a three-layer FBG (Inset: magnified view with the orange bar indicating a length of 20 μm); and (d) the measured transmitted mode profile.

The FBG section was cut and polished using silicon carbide polishing pads to optical quality, in order to expose the fabricated device. A cross-sectional image of the FBG is shown in Fig. 9(c) with a magnified view in the inset. The resulting structure has inner dimensions of ~9 μm and ~8 μm for the major (vertical) and minor (horizontal) axes, respectively. The outer dimensions were ~53 μm and ~26 μm, respectively. Crack formation was again observed both perpendicular and parallel to the FBG directions.

The transmission mode profile of the FBG was measured at 1550 nm in Fig. 9(d) and the FWHM were calculated to be 6.67 μm horizontally and 7.52 μm vertically. The waveguide within the sapphire fiber shows single-mode operation. The noisy mode profile is likely caused by manufacturing defects due to an inhomogeneous fiber shape along the laser writing direction. The curvature of the sapphire varies along its length, but this was not taken into account, leading to some inaccuracy in aberration correction. Changes in the DCW dimension may introduce an increase in propagation loss. This could be improved with real-time feedback of the fiber surface profile to adaptively correct the phase.

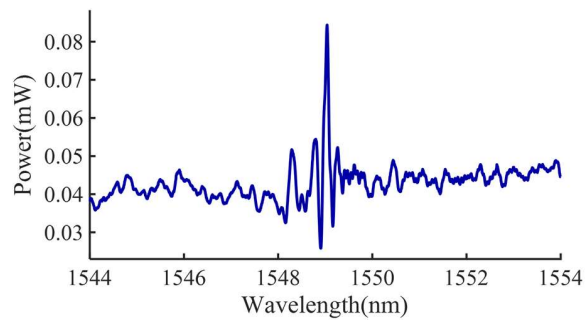

Fig. 10. The experimentally measured reflection spectrum of the single-mode sapphire FBG.

The reflection spectrum of the sapphire FBG was measured, shown in Fig. 10. There is a Bragg resonance, with a center wavelength at 1549.04 nm, corresponding to an effective refractive index of $n_{eff}$= 1.7451. As before, we believe that what appear to be side modes immediately adjacent to the main peak are Fabry-Perot cavity modes, arising from reflection at the end facets. Index matching gel was used on both facets, however as the sapphire has a higher index, residual reflection was inevitable. As previously discussed, this effect could be mitigated with antireflection coatings on the end facets. The bandwidth of the Bragg reflection appears <0.5 nm as a result of the single-mode waveguide. Furthermore, unlike multimode sapphire FBGs, the mode is stable, showing significant promise for accurate measurements in extreme environments.

## 6. CONCLUSIONS

To conclude, sapphire DCWs which are single-mode at telecommunications wavelengths have been demonstrated. The DCWs were fabricated with femtosecond laser direct writing using non-immersion lenses and adaptive optics aberration compensation. We confirmed the DCWs to be single-mode via experimental measurements and simulations. This work was extended to fabricate single-mode sapphire WBGs. This was achieved through a novel ring-shaped geometry incorporating a multi-layer DCW design and cladding-by-cladding inscription. The WBG had a narrow bandwidth (<0.5 nm), and withstood annealing at 1000°C. Finally, we created a single-mode sapphire FBG by writing these structures within 425 μm diameter sapphire optical fiber. This result shows great potential for multipoint quasi-distributed sensing in ultra-extreme environments. Furthermore, single-mode sapphire fiber fabricated with this method enables a wide variety of other challenging sensing and communications applications.

**Funding.** Engineering and Physical Sciences Research Council (EP/T00326X/1, EP/R004803/01)

**Acknowledgments.** The authors gratefully acknowledge the support and advice of their partners Rolls-Royce plc, Cranfield University, UK Atomic Energy Authority and MDA Space and Robotics. The authors thank Tony Wheeler for the use of the furnace. They also thank Professor Dominic O'Brien and Dr Andy Schreier for the use of optical test equipment.

**Disclosures.** The authors declare no conflicts of interest.

**Data availability.** Data underlying the results presented in this paper are available in Dataset 1, Ref. [28].

**Thumbnail Image (200×200 pixels)**

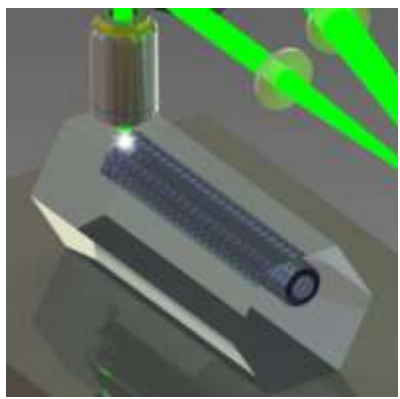